\documentclass[a4paper,12pt]{article}
\usepackage{eurosym}
\usepackage[utf8]{inputenc}
\usepackage{cancel}
\usepackage{ulem}
\usepackage{amsfonts}
\usepackage{amssymb}
\usepackage{graphicx}
\usepackage{amsmath}
\usepackage{enumerate}
\usepackage{mathtools}
\usepackage{subfig}
\usepackage{color}
\usepackage{tikz}
\usepackage{float}
\usepackage{here}
\usepackage{cite}
\usepackage{mathrsfs}
\usepackage{float,epsfig}
\usepackage{dcolumn}
\usepackage[toc,page]{appendix}
\usepackage{graphicx}
\usepackage{bm}
\usepackage{amsmath,amssymb,amsthm}
\usepackage[colorlinks=true,linkcolor=blue,citecolor=red]{hyperref}
\usepackage{multirow}

\usetikzlibrary{calc}
\newcommand{\be}{\begin{equation}}
\newcommand{\ee}{\end{equation}}
\newcommand{\bea}{\setlength\arraycolsep{2pt} \begin{eqnarray}}
\newcommand{\eea}{\end{eqnarray}}

\def\0{{\sst{(0)}}}
\def\1{{\sst{(1)}}}
\def\2{{\sst{(2)}}}
\def\3{{\sst{(3)}}}
\def\4{{\sst{(4)}}}
\def\5{{\sst{(5)}}}
\def\6{{\sst{(6)}}}
\def\7{{\sst{(7)}}}
\def\8{{\sst{(8)}}}
\def\sst#1{{\scriptscriptstyle #1}}

\makeatletter \@addtoreset{equation}{section}

\definecolor{lime}{HTML}{A6CE39}

\tikzset{>=latex}

\begin{document}

\title{\Large \textbf{  On Thermodynamics of  AdS  Black  Holes in  Square-Root Nonlinear Electrodynamics }}
\author{ {\small   A. Belhaj$^1$, S.  E.  Ennadifi$^2$,   M.   Jemri$^1$, M. Nach$^3$\thanks{Corresponding author: maryem.jemri@um5r.ac.ma}
 \footnote{
Authors are listed  in alphabetical order. 
} \hspace*{-8pt}} \\
{\small  $^1$ESMaR, Faculty of Science, Mohammed V University in Rabat, Rabat, Morocco  }\\
{\small  $^2$LHEP-MS,   Faculty of Science, Mohammed V University in Rabat, Rabat, Morocco }
\\
{\small  $^3$LPMS,  Faculty of Science, Ibn Tofail University, Kenitra,  Morocco }
}
\maketitle

\begin{abstract}
Inspired by exponential Maxwell  models,   we propose a new black hole solution in square root nonlinear electrodynamics.  Using the  analogy
between black hole mechanics and thermodynamics,  we investigate  the corresponding   stability, the   $P-v$ criticality and  the phase transitions.  As a result,  we find that the obtained  black hole exhibits  Van der Waals behaviors without additional corrections which typically appear in universal ratio computations.


\end{abstract}

%

\newpage

\section{Introduction}
The study of black holes has become a warm topic in modern physics \cite{1,2,3}. Such activities have been supported and motivated by recent empirical findings provided by the  Event Horizon Telescope (EHT) collaboration showing the first images \cite{4,5,6}. In  this regard, two aspects have been largely investigated being thermodynamics and optics \cite{7,71,8,10,11,111}. The latter offers possibilities for studying the shadow  configurations using light equations of motion.  Shadow curves of non-rotating  and rotating solutions have been  elaborated  for various black hole gravity theories providing interesting results. As a result, many models have been corroborated using constraint parameter scenarios via different approaches including advanced  numerical computations including CUDA parallel programing simulations  and  machine learning methods \cite{12,13,14,15,16,17}. Alternatively, the thermodynamics of the  black holes on the Anti de Sitter space (AdS) has also been  developed by considering the  link between the cosmological constant and the pressure. The  $P-v$ criticality   and certain transition structures have been approached supplying various universal ratios. These  universality   behaviors have been exploited to  support  similarities with Van der Waals fluid systems \cite{29,28,9}.\\
Recently, extended models going beyond ordinary charged  AdS black hole  solutions have been provided using different deformation terms. Mass and charge terms have been introduced in the metric function appearing in the line element of the space-time geometries in which the black holes are built \cite{20}. Concerning charge  additional contributions, many nonlinear electromagnetic field models have been proposed. Such nonlinearity aspects offer possible extensions of the  ordinary Maxwell-Einstein theory by implementing non-trivial functions of field invariants given in terms of the Maxwell tensor and its dual. These functions have provided non-trivial behaviors of black holes in nonlinear electrodynamics (NED). Concretely, the corresponding thermodynamics has been extensively investigated proffering  interesting results in  the nonlinear Maxwell theory coupled to gravity models \cite{21,22,23,24,240,241}.\\
More recently, NED has also been combined  with extended gravity models including the Rham–Gabadadze-Tolley (dRGT) massive gravity to approach black hole thermodynamics \cite{25}. In this context, a NED model has been constructed via an exponential electrodynamics Lagrangian involving the Maxwell field invariants.    Certain thermodynamic properties and phase structures have been studied    showing  a rich thermodynamic behavior   exhibiting  several types of phase transitions in terms of  the magnetic charge values \cite{26}.

Motivated by such activities, we propose a new black hole solution in square root nonlinear electrodynamics (SRNED). Precisely, we  present a SRNED Lagrangian term generating corrections to the ordinary Maxwell invariant fields. In this regard, we investigate the corresponding thermodynamics  by assessing the  stability, the   $P-v$ criticality and  the phase transitions.  As a result,  we find that the obtained  SRNED black hole exhibits  Van der Waals behaviors without additional corrections which typically appear in  the elaboration of universal ratios.

The organization of this work is as follows.  Section 2 exposes  a new black hole solution in  SRNED backgrounds.  Section 3 presents  the associated thermodynamics by computing and analyzing  the relevant quantities.   Section 4  provides    the   $P-v$ criticality and  the phase transitions. The last section  gives concluding remarks and open questions. 
\section{A  novel   black hole solution   in  SRNED}
In this section, we propose a new class of magnetic  black holes in  SRNED backgrounds. Inspired by exponential electrodynamics  reported in \cite{26}, we present a new SRNED term in the Lagrangian computations.  	 To do so, we consider the action describing  the   standard Einstein gravity coupled to NED.  Ignoring the electric charge  sector contributions,  the latter can be rewritten as
\begin{equation}
S=\frac{1}{16\pi}\int d^4x\,\sqrt{-g}
\left[R+4\mathcal{L(F)}\right],
\end{equation}
where $R$ is the Ricci scalar. $\mathcal{L(F)}$ is an arbitrary function of the electromagnetic  field invariant
$
\mathcal{F}=\frac{1}{4}F_{\mu\nu}F^{\mu\nu},
$
where $F_{\mu\nu}=\partial_\mu A_\nu-\partial_\nu A_\mu$ denotes  the electromagnetic field strength tensor. It has been observed that the  NED framework provides richer physical phenomena. This generates a finite self-energy of point charges associated with both regular and non-regular black hole geometries \cite{18,19}. In a nutshell, the corresponding Einstein field equations  are shown to be
\begin{equation}
\left(R_{\mu\nu}-\frac{1}{2}R g_{\mu\nu}\right)
+2g_{\mu\nu}\mathcal{L}
-2\frac{\partial\mathcal{L}}{\partial F}
F_{\mu\gamma}F_{\nu}^{\ \gamma}
=0,
\end{equation}
where $R_{\mu\nu}$ is the Ricci tensor.
For the NED field, the equations of motion  are expressed   as  
\begin{equation}
\nabla_{\mu}\left(
\frac{\partial \mathcal{L(F)}}{\partial \mathcal{F}}
F^{\mu\nu}
\right)=0,\quad
\nabla_{\mu}\left({}^{*}F^{\mu\nu}\right)=0.
\label{14}
\end{equation}
In this work, we consider a static and spherically symmetric metric to approach four-dimensional black hole solutions by focusing  on the  action  describing  SR contributions in NED scenarios.  In this way, the metric line element is assumed to be
\begin{equation}
ds^2=
-f(r)dt^2+\frac{1}{f(r)}dr^2+r^2 d\theta^2+r^2\sin^2\theta \, d\phi^2
\end{equation}
where  $f(r)$ is a radial function to be determined  using SRNED computations. 
To obtain  such  a  radial function, we propose  a new form in  SRNED scenarios
\begin{equation}
 \mathcal{L(F)}
= {\cal F}\sqrt{1-k\left(2q^{2}{\cal F}\right)^{1/4}}
\end{equation}
where $q$ is a constant which could  be considered as a magnetic charge. $k$ is a SRNED  dimensionless  parameter.   To approach the associated  black hole solution, we consider the following magnetic ansatz 
\begin{equation}
F_{\mu\nu}=2\delta^\theta_{[\mu}\delta^\phi_{\nu]}B(r,\theta),
\label{2.37}
\end{equation}
where $B(r,\theta)$ is a function of $r$ and $\theta$, which can be obtained from the NED equations. Calculations provide the following form 
\begin{equation}
F_{\mu\nu}=2\delta^\theta_{[\mu}\delta^\phi_{\nu]}q(r)\sin\theta.
\end{equation}
Using this  electromagnetic  tensor  form, Eq.~(\ref{14}) yields 
\begin{equation}
q(r)=q=\text{constant}.
\end{equation}
Interpreting $q$ as  a magnetic  charge, one has 
\begin{equation}
q=\frac{1}{4\pi}\int_{S_\infty^2}F,\quad
F=\frac{1}{2}F_{\mu\nu}dx^\mu\wedge dx^\nu,
\end{equation}
where \(S_\infty^2\) denotes a two-dimensional real sphere at spatial infinity. To obtain the black hole equations of motion, certain quantities should be calculated.  Considering ${ \cal {F}}=\frac{q^2}{2r^4}$, they are given by
\begin{align}
\mathcal{L(F)}=\frac{q^{2} \sqrt{1-\frac{k  q}{r}}}{2 r^{4}},\qquad 
\frac{\partial \mathcal{L(F)}}{\partial \mathcal{F}}
=
\frac{1-\frac{ 9k q }{8r}}{ \sqrt{1-\frac{k q}{r}}}.
\end{align}
Combining the gravitational and  the Maxwell   field  equations of motion, we get the following two differential equations
\begin{equation}
\frac{f^{\prime}\left(r \right)}{r}+\frac{f \! \left(r \right)-1}{r^{2}}+\frac{q^{2} }{r^{4}}\sqrt{1-\frac{kq}{r}}
 = 0 \label{e13}
\end{equation} 
\begin{equation}
\frac{f^{\prime\prime}(r)}{2}+\frac{f^{\prime}(r)}{r}+\frac{q^{2} }{r^{4}}\sqrt{1-\frac{kq}{r}}-\dfrac{q^2}{r^4}\frac{\left(2-\frac{9 k q }{4r}\right)}{ \sqrt{1-\frac{k q}{r}}}
 = 0.\label{e14}
\end{equation} 
Subtracting Eq.~(\ref{e13}) from Eq.~(\ref{e14}), we find  a relevant   equation of  motion which reads as 
\begin{equation}
\frac{f \! \left(r \right)-1}{r^{2}}-\frac{f^{\prime\prime}(r)}{2}+\frac{q^2}{r^4}\frac{\left(2-\frac{9 k q }{4r}\right)}{ \sqrt{1-\frac{k q}{r}}}=0.\label{ee14}
\end{equation} 
Solving Eq.~(\ref{ee14}), we obtain the following radial metric function
\begin{equation}
f(r)=1+\frac{c_1}{r}+c_2r^{2} -\frac{2 q}{{3 k  r}} \left(1-\frac{k  q}{r}\right)^{\frac{3}{2}},
\end{equation}
where $c_1$ and $c_2$ are integration constants. This metric function generates black hole solutions by fixing such constants. Putting $c_1=-2M$ and $c_2=-\dfrac{\Lambda}{3}$, we get
\begin{equation}
f(r)=1-\frac{2 M}{r}-\frac{2 q}{{3 k  r}} \left(1-\frac{k  q}{r}\right)^{\frac{3}{2}}-\frac{ \Lambda}{3}r^{2}.
\end{equation}
At this level, we would like to  make a few  comments regarding this solution. First, a closer examination shows that this new form  recovers certain known solutions. Taking $q=0$,  for instance, the NED contribution is removed, yielding the  radial metric function 
\begin{equation}
f(r)=1-\frac{2M}{r}-\frac{\Lambda}{3}r^2
\end{equation}
describing  the neutral AdS  black hole solution. Second, the SR term is linked to the NED, controlling the modifications induced by the magnetic charge $q$. Third, taking $kq/r\ll 1$, the SR term can be approximated using a Taylor expansion. This leads to a simplified expression for the metric function
\begin{equation}
f(r)\simeq1-\frac{2M_{eff}}{r}+\dfrac{q^2}{r^2}-\frac{\Lambda}{3}r^2,
\end{equation}
where an effective mass  $M_{eff}$  has emerged 
\begin{equation}
M_{eff}=M+\dfrac{q}{3k}.
\end{equation}
This situation  is distinguishable from that obtained for black holes in exponential NLED. In fact,  the effective mass  term is now  a linear function of $q$, rather than a quadratic one, as found in exponential NED  contributions\cite{26}.

\begin{figure}[!ht]
\centering
\includegraphics[scale=0.55]{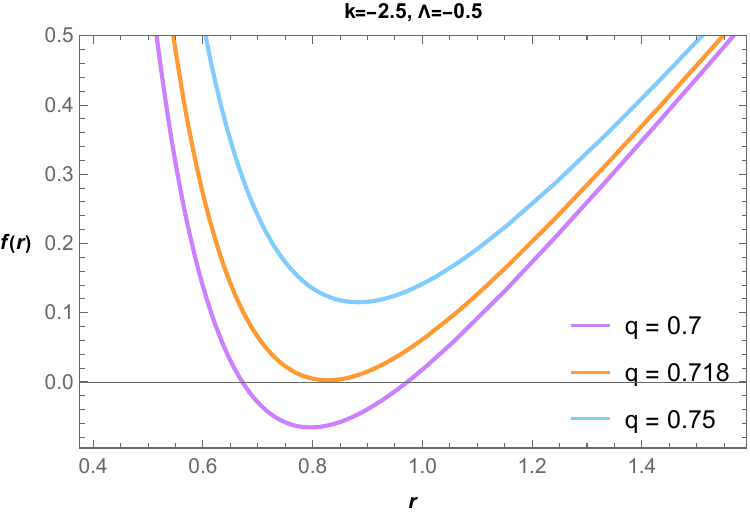} \includegraphics[scale=0.55]{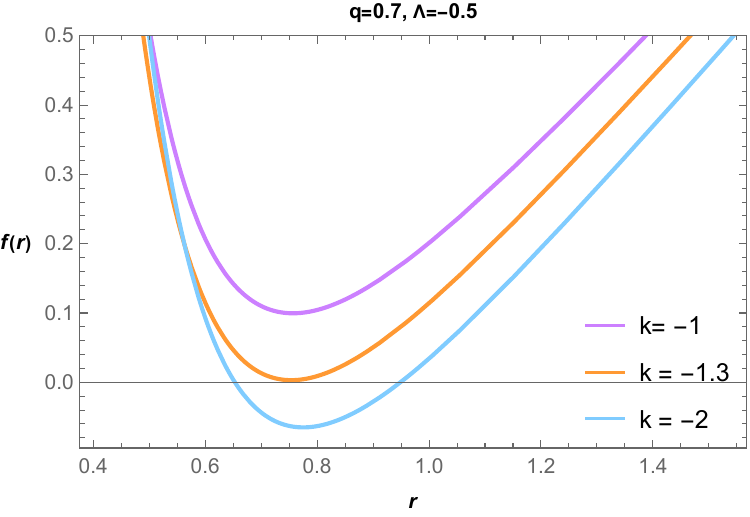}
\caption{\textit{\protect\footnotesize  Effect of $q$ and $k$ on the metric function
$f(r)$}.}
\label{f}
\end{figure}

Before discussing the thermodynamic properties of such a new solution, we
first examine the black hole metric function behaviors. Fixing the mass and the cosmological
constant, the discussion will be elaborated in terms of two relevant parameters being $q$ and  $k$.
Roughly, Fig.(\ref{f}) illustrates such behaviors.
The metric function $f(r)$ exhibits three configurations depending on the magnetic  charge $q$ and the NED parameter $k$. The first case corresponds to a non-extremal black hole with two distinct horizons for $q<q_c$. The second is associated with  an extremal black hole with a double horizon for $q=q_c$, and the third   one describes  a naked singularity for $q>q_c$. Increasing  the magnetic charge  $q$, the two horizons approach each other and eventually merge. Additionally,  it  has been also observed that the NED parameter $k$ has an opposite effect to that of the magnetic  charge $q$.

Having discussed the  construction of such  a  SRNED black hole, we move now to approach the associated thermodynamic aspects  by considering  the analogy
between  black hole physics  and thermodynamics. 
\section{ Black hole stability   in   SRNED backgrounds }
In this section, we study the local and  the global stabilities of the obtained  charged black hole in SRNED. To approach such behaviors, certain quantities should be computed including   the temperature,  the heat capacity, and  the Gibbs free energy. 

\subsection{ SRNED  thermal behaviors}
\begin{figure}[h!]
    \centering
    \begin{tabular}{cc}
       \includegraphics[width=7.5cm,height=5.5cm]{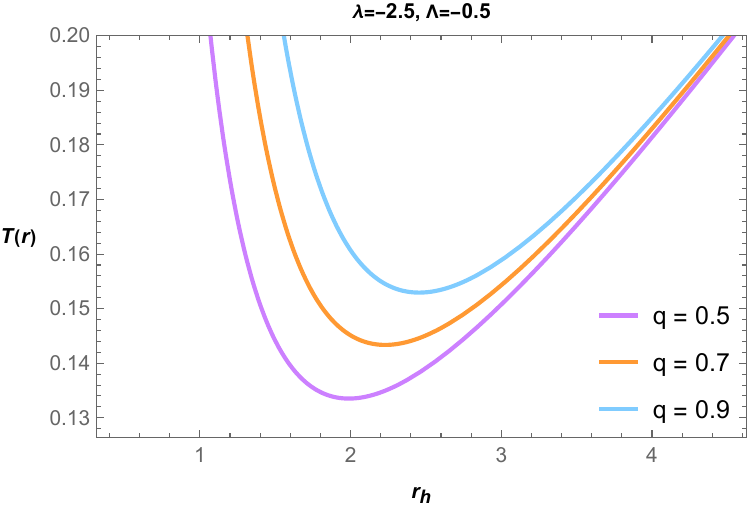} & 
         \includegraphics[width=7.5cm,height=5.5cm]{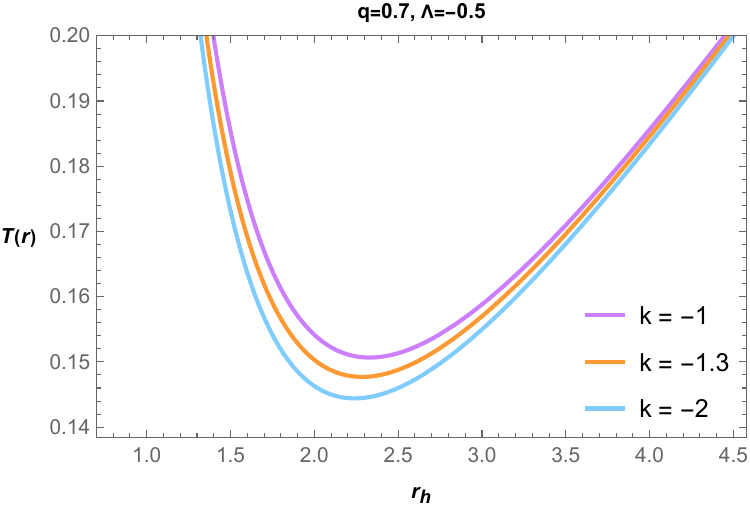} 
    \end{tabular}
    \caption{ Hawking temperature as a function of the horizon radius for different values of $q$ and $k$.}
    \label{Fig3.2}
\end{figure}

To approach the thermodynamics of such  a  SRNED  black hole, we  need to   calculate  the relevant   quantities such  as  the mass and the temperature.  To  obtain the corresponding expressions,   we  first compute    the mass quantity being a function of the horizon radius $ r_h $. This can be obtained by  solving the constraint $ f(r_h) = 0 $.  Concisely,   it is shown to be
\begin{equation}
M = \frac{r_h}{2}-\frac{q \left(1-\frac{k q}{r_h}\right)^{ \frac{3}{2}}}{3 k \,r_h^{2}}-\frac{ \Lambda}{6} r_h^{3}.    
\end{equation}
To calculate  the  Hawking temperature, we use the relation  $T_H = \kappa / (2\pi) $, where $ \kappa $ is the surface gravity given by
\begin{equation}
\kappa = \left. \frac{ \mathrm{d}f(r) }{ \mathrm{d}r } \right|_{r = r_h}.    
\end{equation}
After computations, the  Hawking temperature is found to be
\begin{equation}
T_H =\frac{1-\Lambda  \,r_h^{2}}{4 \pi r_h }+\frac{q \! \left(7 k q  -4 r_h \right) }{12 \pi  k \,r_h^{3}}\sqrt{1-\frac{k q}{r_h}}.  \label{T} 
\end{equation}
It turns out that certain know expressions could be recovered by taking appropriate limits. Considering $ q= \Lambda=0$,   for instance, we get  the Schwarzschild black
hole temperature  given by $T_{H}=\frac{1}{4\pi r_h}$ \cite{27}.
To examine the thermal variation, we illustrate the temperature as a function of $r_h$ in Fig.~(\ref{Fig3.2}).
Increasing the magnetic  charge $q$, the temperature curves
show a more pronounced decline, and the highest minimum value corresponds to the lowest charge. At
 certain values of $r_h$, all the temperature curves converge, indicating a universal behavior
independent of the magnetic  charge $q$. Increasing the NED  parameter $k$, we observe an opposite behavior, with the highest minimum corresponding to the largest value of the SRNED parameter  $k$.
\subsection{Stability behaviors} 
\begin{figure}[h!]
    \centering
    \begin{tabular}{cc}
    \includegraphics[width=8cm,height=6cm]{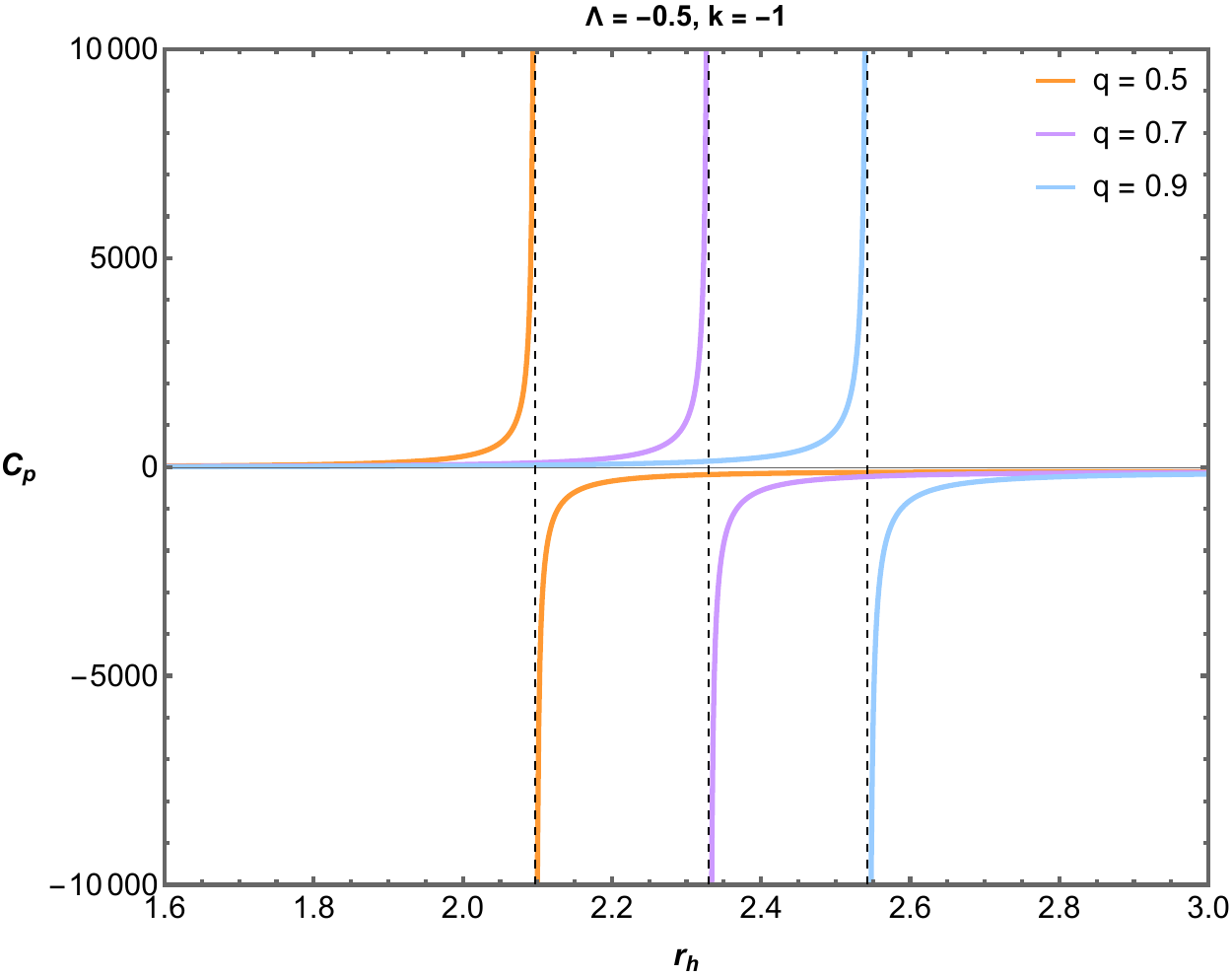} 
    \includegraphics[width=8cm,height=6cm]{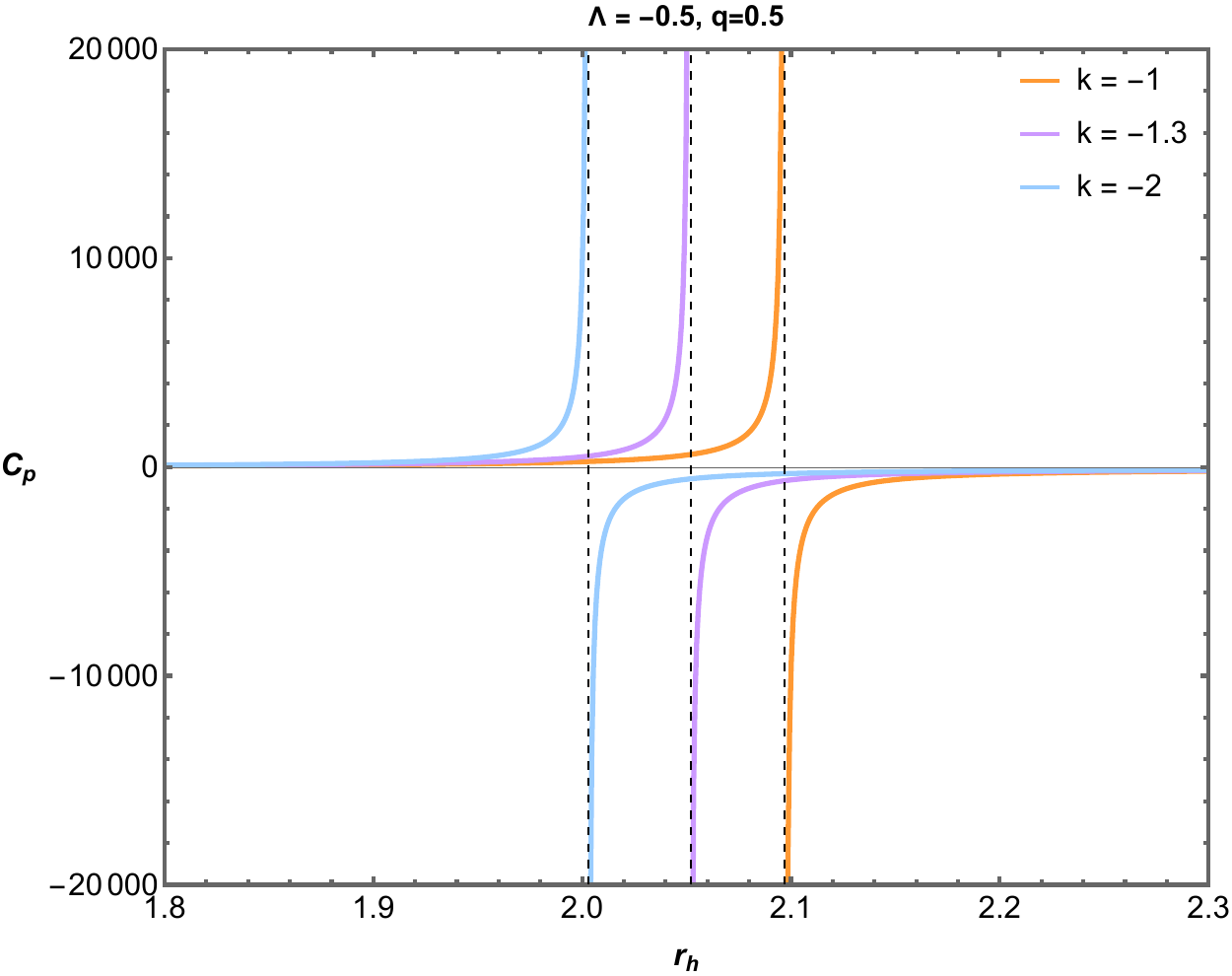} 
    \end{tabular}
    \caption{Local stability behaviors.}
    \label{4}
\end{figure}

To discuss the local  stability of such black holes, we need to compute  the heat capacity via the following
relation
\begin{equation}
C_p = T_H  \frac{\partial S}{\partial T_H}.
\end{equation}
To calculate such a quantity, one should first determine the entropy using the Bekenstein–
Hawking area law $S = \dfrac{\mathcal{A}}{2}=\pi r_h^2$,  where $\mathcal{A}$  is the surface area of the black hole event horizon given by  $\mathcal{A} =\int\int\sqrt{g_{\theta \theta}g_{\phi \phi}}=2 \pi r_h^2.$
 Using the Hawking temperature expression, the heat capacity is found to be
\begin{equation}
C_p =\frac{12 \pi  k r_{h}^{2} \left(r_{h}^{3}\left(2 \Lambda -1\right) \sqrt{1-\frac{k q}{r_{h}}}\, +q^{2} \left(r_{h}-k q \right)\right)}{6 k r_{h}^{3} \left(\Lambda  r_{h}^{2}+1\right) \sqrt{1-\frac{k q}{r_{h}}}-q \! \left(2  \left(8-\dfrac{31 k q}{r_{h}}  \right)+49  k^{2} q^{2}\right)}.
\end{equation}
Taking $ q= \Lambda=0$, for example,   we recover the  expression relative to the Schwarzschild black
hole   given by $
C_p =-2  \pi r_h^{2} $\cite{27}.
According to   the sign of the thermal capacity, we can approach the stability of the obtained black hole solution  in SRNED. Indeed, a locally stable thermodynamic system can occur if $C_{p}>0$, while an unstable solution occurs if $C_{p}<0$,  In Fig.~(\ref{4}), we illustrate $ C_{p} $ as a function of $r_{h} $ for selected result in the moduli space.
For a general point in  a  such space, we observe  that the heat capacity curves are discontinuous at the critical values $ r_{h} = r_{h}^{c} $. At these points, the heat capacity $ C_p $ exhibits divergent behaviors. This indicates a second-order phase transition.

\begin{figure}[h!]
    \centering
    \begin{tabular}{cc}
    \includegraphics[width=8cm,height=6cm]{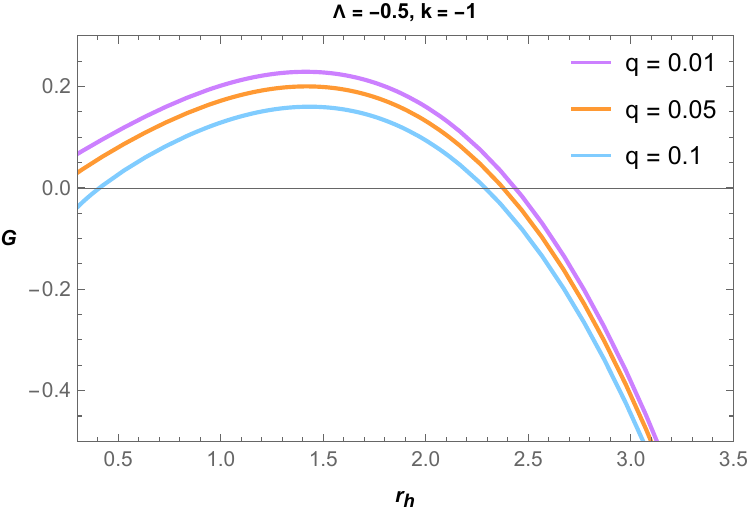} 
    \includegraphics[width=8cm,height=6cm]{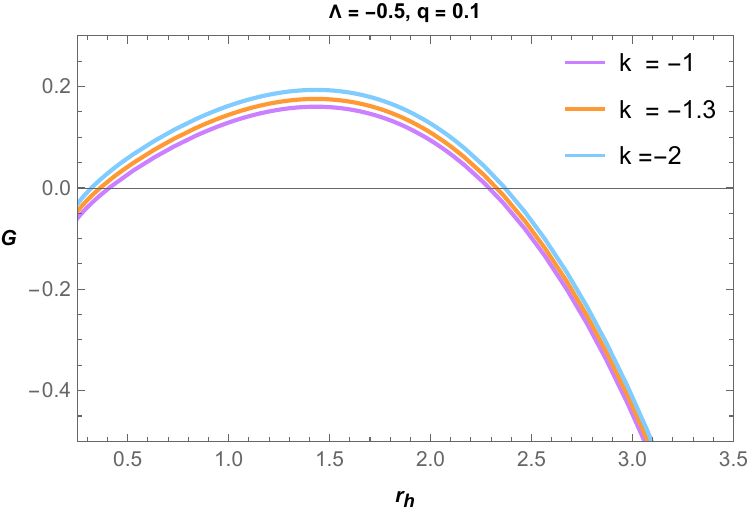} 
    \end{tabular}
    \caption{Global stability behaviors.}
    \label{G}
\end{figure}

Now, we examine  the global thermodynamic stability by approaching  the Gibbs free
energy,  through the relation 
\begin{equation}
G=M-TS.
\end{equation}
The computations reveal that the Gibbs free energy takes the following form 
\begin{equation}
G=\frac{r_{h} \left(\Lambda  r_{h}^{2}+3\right)}{12}-\frac{q\left(11 k q -8 r_{h}\right)}{12 k r_{h}} \sqrt{1-\frac{k q}{r_{h}}}.
\end{equation}
Removing the magnetic charge,  the Gibbs free energy
reduces to
\begin{equation}
G=\frac{1}{4} r_h\left( 1+\frac{r_h^{2} \Lambda }{3}  \right) ,
\end{equation}
recovering the Schwarzschild black hole  expression \cite{272}. A black hole is globally stable when its Gibbs free energy is negative ($G<0$), whereas a positive value ($G>0$) indicates that the black hole is not globally favored. The condition $G=0$ represents a change in the sign of the Gibbs free energy rather than a genuine phase transition, such as the Hawking--Page transition, since no thermal AdS reference background is considered. Fig.~(\ref{G}) shows the variation of $G$ as a function of the horizon radius $r_H$ for physically acceptable parameter values.
\section{$P-v$ criticality and phase transition   in SRNED  magnetic black holes }
In this section, we investigate the P-V criticality,  the Joule–Thomson expansion,  and the  phase transitions of  such  a  black hole in SRNED backgrounds. To start, we need to  establish the  thermodynamic equation of state. In the extended phase space,   this equation can elaborated by considering  the cosmological constant $\Lambda$   as a thermodynamic variable \begin{equation}
P = -\frac{\Lambda}{8\pi}.
\end{equation}
This link  not only provides a more complete thermodynamic description, but also favors the emergence of non-trivial  phase structures and critical phenomena similar to those encountered in classical thermodynamic systems, such as Van der Waals fluids.

\subsection{$P-v$   universal behaviors}
\begin{figure}[h!]
    \centering
       \includegraphics[width=10cm,height=7cm]{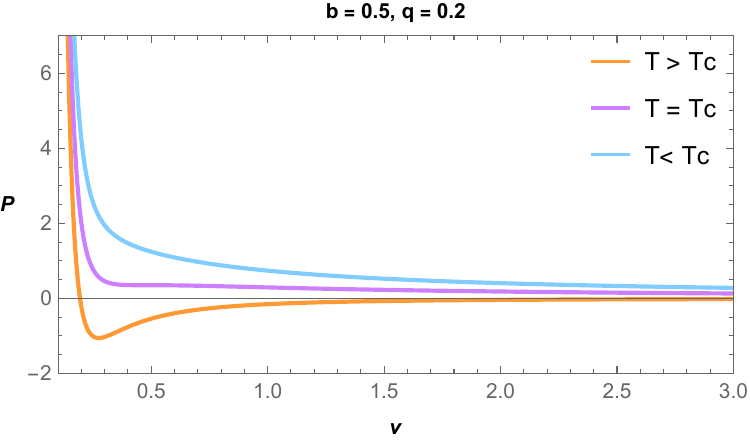} 
\caption{Pressure in terms of $v$ for different values of $T$ with $q=0.2$ and $b=0.5$. }
\label{F41}
\end{figure}

 This  interplay is the starting point for determining the equation of state needed to check  the $P-v$ criticality of the proposed black hole  system \cite{28}. Using the temperature and  the pressure expressions, we find the following state equation
\begin{equation}
 P=\frac{4 \pi  T r_{h}-1}{8 \pi r_{h}^{2}}-\frac{q \left(7 k q -4 r_{h}\right) }{24 \pi  k r_{h}^{4} }\sqrt{1-\frac{k q}{r_{h}}}.\label{P}
\end{equation}
Vanishing the charge,  $q=0$, we recover the expression   of the ordinary neutral black hole 
\begin{equation}
 P =\frac{4 \pi  T r_{h}-1}{8 \pi r_{h}^{2} }.
\end{equation}
To determine the thermodynamic critical  values,  the volume quantity  is needed. Indeed, it reads as 
 \begin{equation}
 V=\frac{4 \pi  \,r_h^{3}}{3}.\label{V}
 \end{equation}
A close examination reveal that the determination  of the critical quantities is  a tough task.  However,  we could follow techniques   explored in  \cite{291,292}.   Introducing a new parameter  $b$ expressed as follows\begin{equation} 
b=\frac{2 q k}{v}, 
\end{equation}
where $v$ is the  specific volume, the  thermodynamic critical values could be determined by imposing appropriate conditions. Indeed, the  critical values  can be determined by solving the following constraints  \begin{equation}
\frac{\partial P}{\partial v }=0,\hspace{1.5cm}\frac{\partial ^{2}P}{%
\partial v^{2}}=0. 
\end{equation}
After computations, the critical thermodynamic  quantities are found to be 
\begin{align}
P_c =& \frac{\sqrt{2}\, b \sqrt{2-2 b }}{64 \pi q^{2}  \left(7 b -4\right) \left(b -1\right)},  \\ 
T_{c} =& \frac{b 2^{{3}/{4}}}{6\, \pi q \sqrt{b\left(4  -7 b\right) \sqrt{2-2 b }}}, \\
v_{c} =& \frac{2 \,2^{{1}/{4}}\, q \sqrt{b\left(4-7 b \right) \sqrt{2-2 b }}}{b}.
\end{align}
This  critical triple $(P_{c}, T_{c}, v_{c})$ generates directly  the  famous ratio
\begin{equation}
\chi = \frac{P_c v_c}{T_c} = \frac{3}{8}, 
\end{equation}
without extra contributions   which usually  depend on  the modification  parameters.  This demonstrates that the obtained results are in excellent agreement with the universal Van der Waals behaviors governed by the charge $q$ and the SRNED parameter $k$ \cite{29}. To support these Van der Waals behaviors,  we present the   $P$-$v$
diagram, as illustrate in Fig.~(\ref{F41}).

At temperature $T$ above the critical value $T_c$, the black hole behaves like an ideal gas. The critical isotherm at $T = T_c$ is  shown  by an inflection point at the critical volume $v_c$ and the critical pressure $P_c$. At $T < T_c$, there is an unstable thermodynamic region.  It  has been remarked that the $P$-$v$ diagram clearly resembles that of a Van der Waals fluid system.

\subsection{Phase transitions}
\begin{figure}[h!]
    \centering
     \includegraphics[width=7cm,height=6cm]{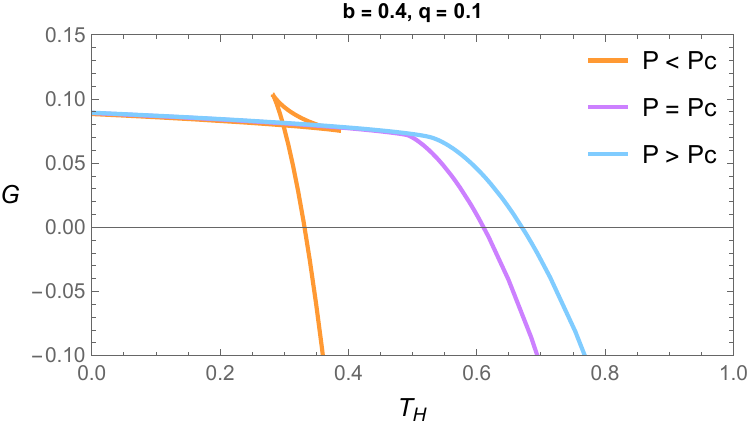} 
    \caption{Gibbs free energy in terms of the temperature for different values of $P$.}
    \label{7}
\end{figure}

Using the $P-v$ criticality results, we study the phase transitions by approaching the Gibbs free energy as a function of the  pressure.  Roughly, it  is  proved as
\begin{equation}
G = \frac{r_{h} \left(3-8 \pi P   r_{h}^{2}\right)}{12}+\frac{q \left(8 r_{h}-11 k q \right) }{12 k r_{h}}\sqrt{1-\frac{k q}{r_{h}}}.
\end{equation}
Using  the critical   thermodynamic   values,  the associated behaviors are presented in Fig.~(\ref{7}).
The $G-T$ curves, describing  the Gibbs free energy as a function of temperature, involve similar shapes for different values of the critical pressure $P_c$. 
For pressures below the critical value ($P < P_c$), the diagram reveals a swallow-tail shape. 
This shape signals a first-order phase transition between small and large black holes. 
This behavior is very similar to Van der Waals fluids, supporting the analogy between  such SRNED  black holes and classical fluids.

\subsection{Joule–Thomson expansion}
To unveil more thermodynamic information of   the SRNED  gravity model, we discuss the Joule–Thomson expansion \cite{30,31,32}. Fixing the charge, the Joule-Thomson coefficient can be written as
\begin{equation}
\mu=\left( \dfrac{\partial T}{\partial P} \right)_{M}=\dfrac{1}{C_{P}} \left[ T \left( \dfrac{\partial V}{\partial T} \right)_{P}-V  \right]. \label{m}
\end{equation}
To get the associated expression, we approach  the equation of state  in  terms of the thermodynamic volume.
Using Eq.~(\ref{V}), Eq.~(\ref{P}) and Eq.~(\ref{T}),   the temperature as a function  of the volume and the pressure is found to be 
\begin{equation}
T = \frac{ \! \left(7 b -4\right)q^{2} \sqrt{1-b}}{9 V b}+P \! \left(\frac{6 V}{\pi}\right)^{{1}/{3}}+ \frac{1}{12V} \left(\frac{6V}{\pi}\right)^{{2}/{3}}. \label{t}
\end{equation}
 Eq.~(\ref{t}) and the second part of Eq.~(\ref{m}) give the temperature corresponding  to a zero Joule-Thomson coefficient.  The calculations show  that  the   repeated inversion temperature $T_i$ is
\begin{equation}
T_i = \frac{\left(4-7 b \right) q^{2}   \sqrt{1-b}}{9 V b}+\frac{P }{3}\left(\frac{6V}{\pi}\right)^{{1}/{3}}-\frac{ 1}{36 V}\left(\frac{6V}{\pi}\right)^{{2}/{3}}.  \label{tt}
\end{equation}
Using  the volume quantity,  this temperature   is shown  to be  \begin{equation}
T_i= \frac{\left(4-7 b \right) q^{2} \sqrt{1-b }-b \, r_i^{2} \! \left(8 \pi P_i   \,r_i^{2}-1\right) }{12 \pi b \,r_i^{3} } .
\label{ti1}
\end{equation}
Exploiting  Eq. (\ref{t}),  we obtain
\begin{equation}
T= \frac{\left(7 b -4\right) q^{2} \sqrt{1-b }+3\, b\, r^{2}  \! \left(8 \pi P   \,r^{2}+1\right)}{12  \pi\, b \, r^{3} }.\label{ti2}
\end{equation}
Subtracting Eq.~(\ref{ti1}) form Eq.~(\ref{ti2}),  we  set an algebraic equation which reads as
 \begin{equation}
\left(7 b-4 \right) q^{2} \sqrt{1-b }+2 \, b\, r_i^{2} \! \left(4 \pi  P_i  \,r_i^{2}+1\right)= 0, 
 \end{equation}
where $P_{i}$  indicates the inversion pressure. Considering only the positive root, we find
 \begin{equation}
r_i = \frac{\sqrt{2}}{4 \sqrt{\pi}\, P_i b}\, \sqrt{-P_i b \left(b +\sqrt{\left(8\,\pi  P_i \,q^{2}  \left(4-7 b \right) \sqrt{1-b }-b \right) b}\right)}.
\end{equation}
By inserting this root into Eq.~(\ref{ti1}), we obtain the inversion temperature  
\begin{small}
\begin{equation}
T_{i}=\frac{\sqrt{2}\, b P_i \left(3 \sqrt{b \left(b -8 \, \pi  \,P_i \,q^{2} \left(7 b -4\right) \sqrt{1-b }\right)}+16 \pi  P_i \,q^{2} \left(7 b -4\right) \sqrt{1-b }-3 b \right)}{6 \sqrt{\pi}\, \sqrt{P_i b \left(-b +\sqrt{b \left(b -8 \pi  P_i \,q^{2} \left(7 b -4\right) \sqrt{1-b }\right)}\right)}\, \left(b -\sqrt{b \left(b -8 \pi  P_i \,q^{2} \left(7 b -4\right) \sqrt{1-b }\right)}\right)}.
\end{equation}
\end{small}
 At zero inversion pressure $P_i=0$, the inversion temperature reaches  its minimum value
 \begin{equation}
T_{i}^{min}=\frac{b \sqrt{2}}{12\, \pi \, q  \sqrt{b \sqrt{1-b }\, \left(4-7 b \right)}}. 
 \end{equation}
The  relationship between the minimum inversion and the  critical temperatures  generates a  ratio 
 \begin{equation}
\xi= \dfrac{T_{i}^{min}}{T_{c}}.
 \end{equation} 
The computations associated with    such  a  SRNED  black hole provides 
  \begin{equation}
\xi=\dfrac{1}{2}.
 \end{equation} 
This reveals that the obtained result matches perfectly the universal behaviors of charged AdS black holes with respect to  $q$ and  $k$ \cite{33}. The $P-v$ criticality and  the Joule–Thomson expansion universal behaviors support the validity of the proposed black hole metric in  SRNED  scenarios. This encourages the implementation  of alternative investigations  to unveil more physical aspects of SRNED backgrounds.

\section{Conclusion and open questions}

In this work, we have presented a novel class of charged AdS black hole solutions coupled to an NED model   via  a  SR Lagrangian term  given in terms of the Maxwell field invariant $\cal F$.  To achieve such a  solution, we have first calculated certain  geometric quantities and substituted them into Einstein field equations. These computations have generated an equation of motion  solved by  the black hole metric function of SRNED backgrounds . Subsequently, we have investigated the thermodynamic properties by computing the relevant quantities.  Precisely,  we have examined  the thermal stability and the criticality behaviors, including  the Joule–Thomson expansion and  the phase transitions. To analyze the thermal stability, we have computed the heat capacity and the  Gibbs free energy. Additionally,   we have explored the $P-v$ criticality  behavior by computing the associated critical points.  Computing the critical thermodynamic values,  we have shown that the ratio $\dfrac{P_cv_c}{T_c}$  is a universal quantity which does not depend on the NED contributions. In fact, it  has been shown that it  recovers  directly the behavior of a Van der Waals fluid without additional corrections which could be removed by considering appropriate regions in the moduli space. Then, we have examined the phase transition using $P-v$ criticality results. Finally, we have investigated the Joule–Thomson expansion for  such a  black hole solution showing  similarities with  the four-dimensional Reissner-Nordstrom-AdS  black hole. The universal behaviors  support  the validity of the proposed black hole metric in SRNED  scenarios.

This work  raises certain open questions. A natural question is to consider other properties
including the optical ones.   Bridging  the non-rotating and rotating  shadow curves  with EHT  empirical findings will be elaborated in \cite{ex}.   Other physical aspects of such black holes could be cheeked using different methods including machine learning  approaches  with  appropriate neural networks to approach non-trivial gravity extensions. 

\section*{Acknowledgements}
The authors (AB and MJ)  would like to thank S. E. Baddis and H. Belmahi for recent collaborations on related topics.  MJ gratefully acknowledges the financial support of the CNRST in the
frame of the PhD Associate Scholarship Program PASS.

\end{document}